# At What Distance Can the Human Eye Detect a Candle Flame?*

Kevin Krisciunas and Don Carona
Texas A&M University
Department of Physics and Astronomy
4242 TAMU, College Station, TX 77843

*Abstract* - Using CCD observations of a candle flame situated at a distance of 338 m and calibrated with observations of Vega, we show that a candle flame situated at ~2.6 km (1.6 miles) is comparable in brightness to a 6th magnitude star with the spectral energy distribution of Vega. The human eye cannot detect a candle flame at 10 miles or further, as some statements on the web suggest.

A recent Centrum Silver TV ad for vitamins, narrated reassuringly by Martin Sheen, claims that the human eye can detect a candle flame at a distance of 10 miles.[1] Web searches on the question posed in our title suggest that the correct answer might be 3 miles, or as far as 30 miles! Clearly, we can do better by considering it as a problem of astronomical detectability. Some data would be a considerable help too.

Let us start with a slightly different question. At what distance would a candle flame be comparable to the brightest stars in the sky? Let's call that a star of apparent magnitude 0, such as Vega. Experiments by one of us (KK) and various students indicate that the distance is greater than 150 yards, possibly as far as 400 yards.

We carried out observations at Texas A&M's Physics Observatory using an SBIG uncooled CCD camera of 35 mm aperture and focal length 100 mm (see Fig. 1). The pixel size of this camera is 7.4 arc seconds, the gain is 0.5 electrons/ADU, and the read noise is 8.6 electrons. The quantum efficiency (QE) of the camera (provided by the manufacturer) is shown in Fig. 2. As one can see, the QE peaks at roughly 480 nm (4800 A). In our experience very little starlight blueward of 370 nm is transmitted at our low elevation site.

On 4 and 6 October 2014 (UT) we took sets of 5 exposures of Vega under clear sky conditions, with *nominal* exposure times of 1 to 10 milliseconds. Analysis of the number of counts vs. exposure time (Fig. 3) reveals two things. The true exposure times are about 1.7 milliseconds longer than the nominal exposure times. Otherwise, the counts would not be zero at zero exposure. Also, there is some raggedness in the data owing to scintillation in the Earth's atmosphere. This is a consequence of having such a small aperture.

*Based on a poster presented at the Seattle AAS meeting in January, 2015.

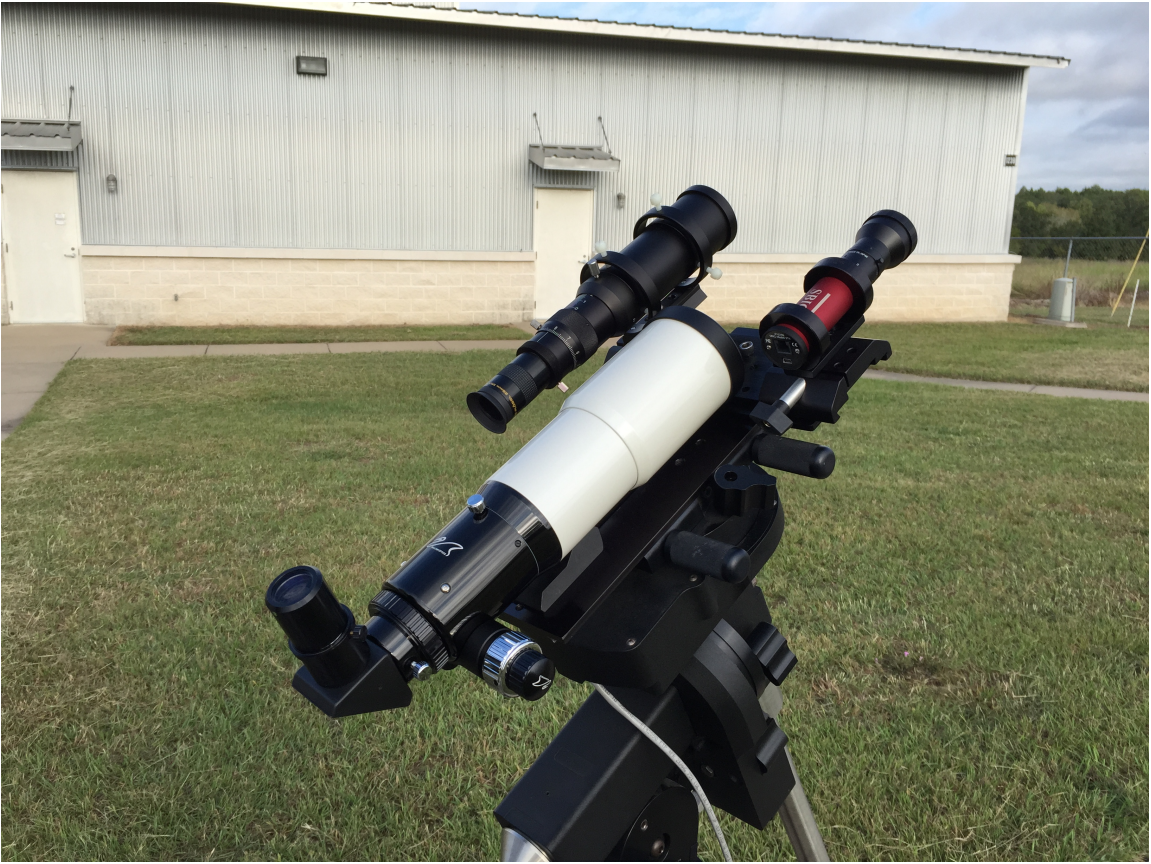

Fig. 1 – The SBIG camera is the small red tube on the upper right. The other two telescopes are just finder scopes.

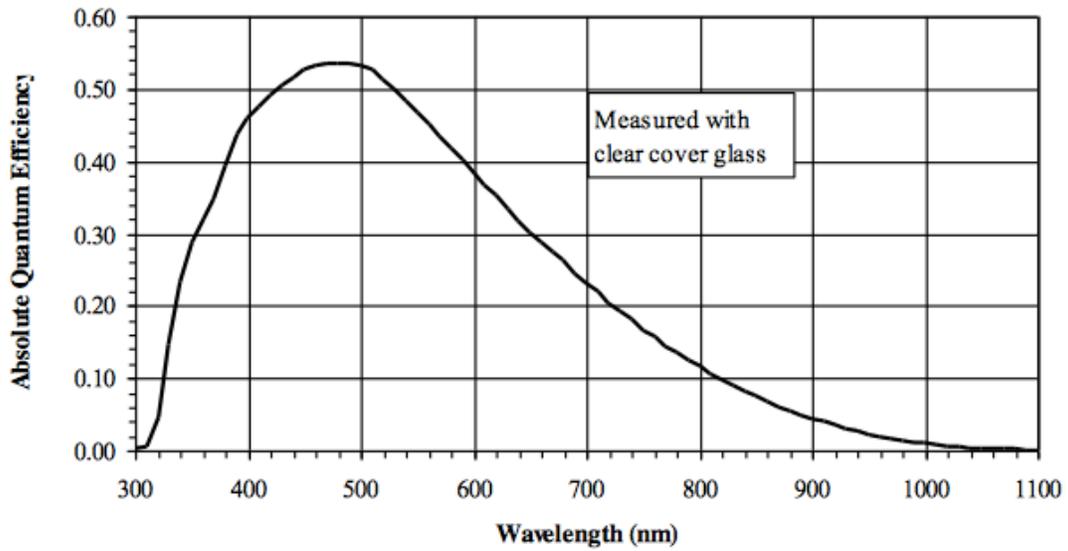

Fig. 2 – Quantum efficiency of SBIG camera as a function of wavelength.

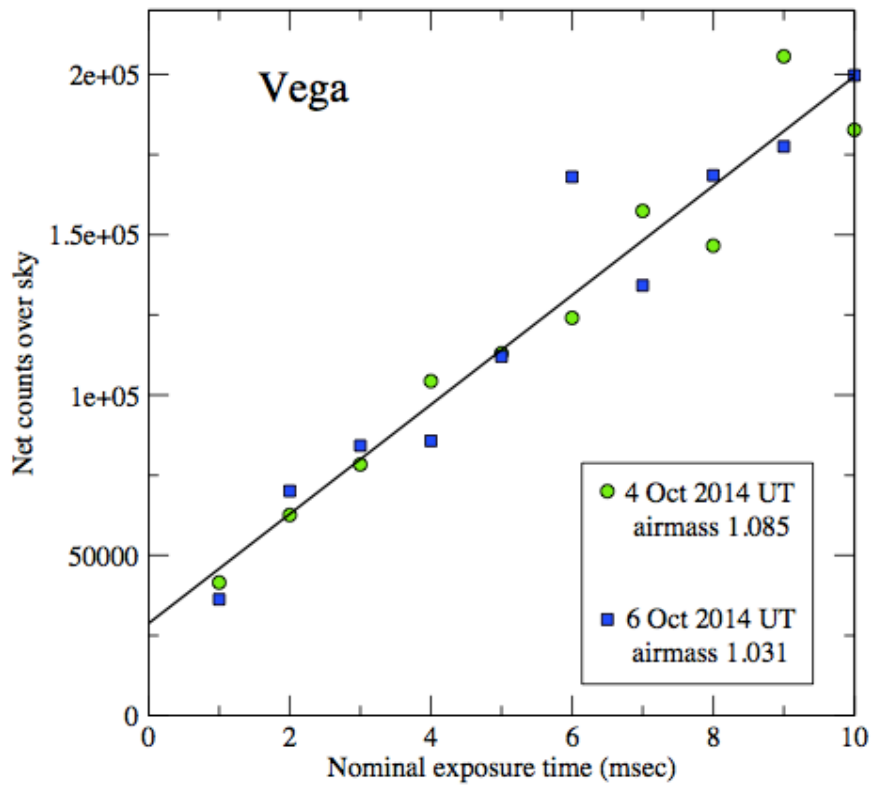

Fig. 3 – Net counts over sky vs. exposure time. We plot averages of 5 integrations per point.

On 6 October 2014 (UT) we also set up a candle flame at a distance of 338 m. To our eyes the candle flame and Vega appeared of comparable brightness, but we found that we saturated the CCD for nominal exposures of 6 milliseconds and longer

We obtained aperture magnitudes of Vega and the candle flame in the IRAF environment,[2] using the apphot package. The instrumental magnitudes of Vega do not have to be corrected to out-of-atmosphere values by correcting for the effects of atmospheric extinction. We needed a very large software aperture (r = 18 px) to include light of the candle flame in the wings of the profile. This was a surprise. The candle flame at maximum brightness was 25 mm high and 7 mm at its base. An object 25 mm in size at 338 m subtends an angle of 15.3 arc seconds, or just over 2 pixels in our camera. We surmise that the light in the wings of the profile is due to scattering of the candlelight by the air surrounding it.

Correcting for the systematic error in the exposure times and combining all the Vega observations on the two nights, we obtain a mean zeropoint of 6.932 +/- 0.026 mag. The candle flame observed with nominal exposure times of 1 to 5 milliseconds gives a corresponding instrumental magnitude of 4.509 +/- 0.054. Thus, the candle flame at 338 m, as measured by the CCD camera, was 2.423 mag brighter than Vega, even though they looked comparable in brightness to our eyes. Since each magnitude corresponds to a factor of the fifth root of 100 (~2.51189) in brightness, the CCD camera obtained a signal from the candle flame at 338 m that was $2.51189^{2.423}$ = 9.315 times brighter than Vega. This is the fundamental observational fact of our CCD integrations on Vega and a candle flame.

To understand this, we need to consider the spectral energy distribution of a black body[3] as a function of wavelength:

$$u_\lambda(T) = \frac{8\pi hc}{\lambda^5} \frac{1}{e^{hc/\lambda k_B T} - 1}$$

We want to consider the number of *photons* we detect from two black bodies (Vega and a candle flame), so we divide $u$ by the energy of the photons ($hc/\lambda$) and use a simple computer program to calculate this every 10 A for a range of wavelength $\lambda$.

As Vega's spectral type is A0 V, we can approximate its spectral energy distribution as a black body with a temperature of 10,000 deg K. This is not strictly true, as hydrogen absorption lines diminish the integrated flux of Vega at the blue end of the optical range of wavelengths. The temperature of a candle flame is roughly 1400 deg K. The reader will know that the total integrated flux of a black body (measured in ergs/cm$^2$/sec) increases proportional to the fourth power of the temperature (the Stefan-Boltzmann law).

Fig. 4 shows two loci. One is the photon number density of Vega multiplied by the SBIG QE curve. The other is the corresponding photon number density from a 1400 deg K black body, multiplied by the SBIG QE curve and scaled by 6.016 X 10$^6$, a scale factor needed to produce the *observed* ratio of signals from Vega and the candle flame.

Next we must consider the luminosity functions of the human eye,[4] basically the filter response of our eye and retina (Fig. 5). For daytime (photopic) vision we may approximate the response of the eye and retina to peak at 5550 A, with a Gaussian half-width of 500 A. The response of the eye and retina under low light conditions is a different "filter" that peaks at 5070 A, and is slightly asymmetrical.

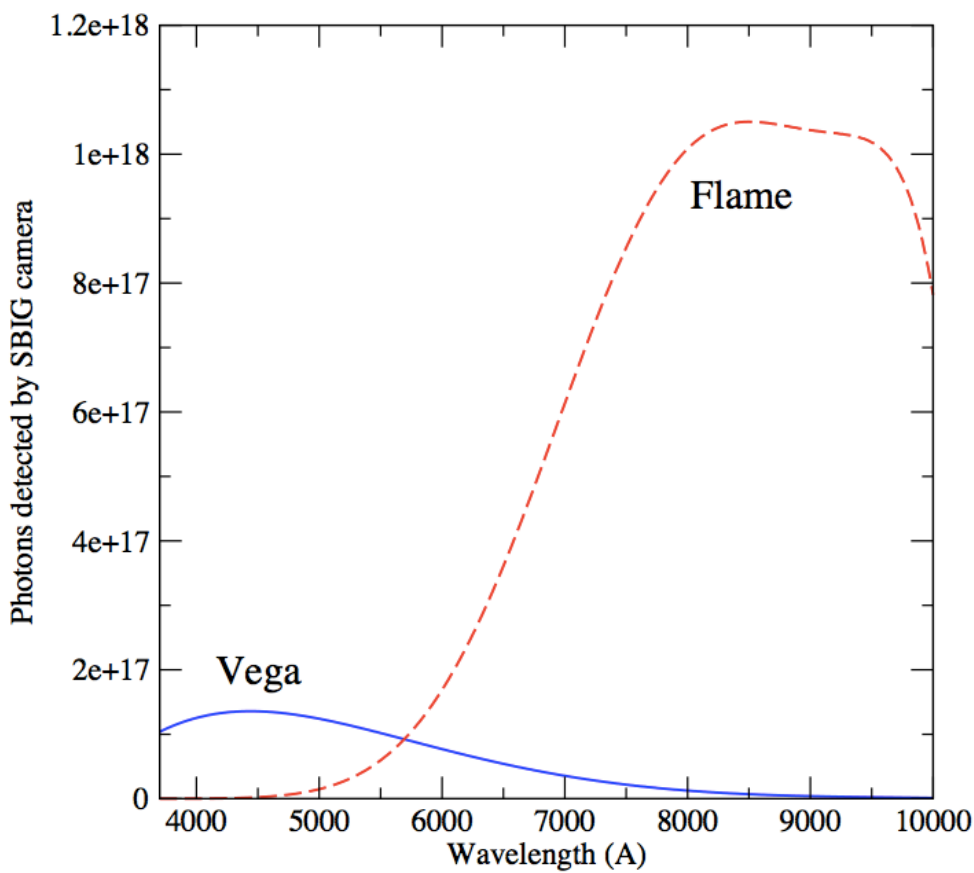

Fig. 4 – Photon number density for Vega (10,000 deg K black body, from Planck's law), multiplied by the SBIG QE curve, and for candle flame (1400 deg K black body), multiplied by the SBIG QE curve and scaled by 6.016 X $10^6$ to give observed ratio of signals of 9.315, as derived from CCD observations.

Returning to the question of the distance of the candle flame such that it would appear equal to Vega in brightness according to our eyes, we take the photon number density of a 10,000 deg K black body vs. wavelength, multiply it by the phototopic luminosity curve of the eye, and integrate this from 3700 to 10,000 A. For the candle flame we take the photon number density of a 1400 deg K black body, multiply it by the photopic luminosity curve of the eye, scale it by 6.016 X $10^6$, and integrate from 3700 to 10,000 A.  The ratio (flame divided by Vega) is ~1.344.  (See top half of Fig. 6)  Since light intensity decreases proportional to $1/d^2$, if the candle flame had been at 338 m times the square root of 1.334, or 392 m, it would have exactly matched the brightness of Vega to our eyes.

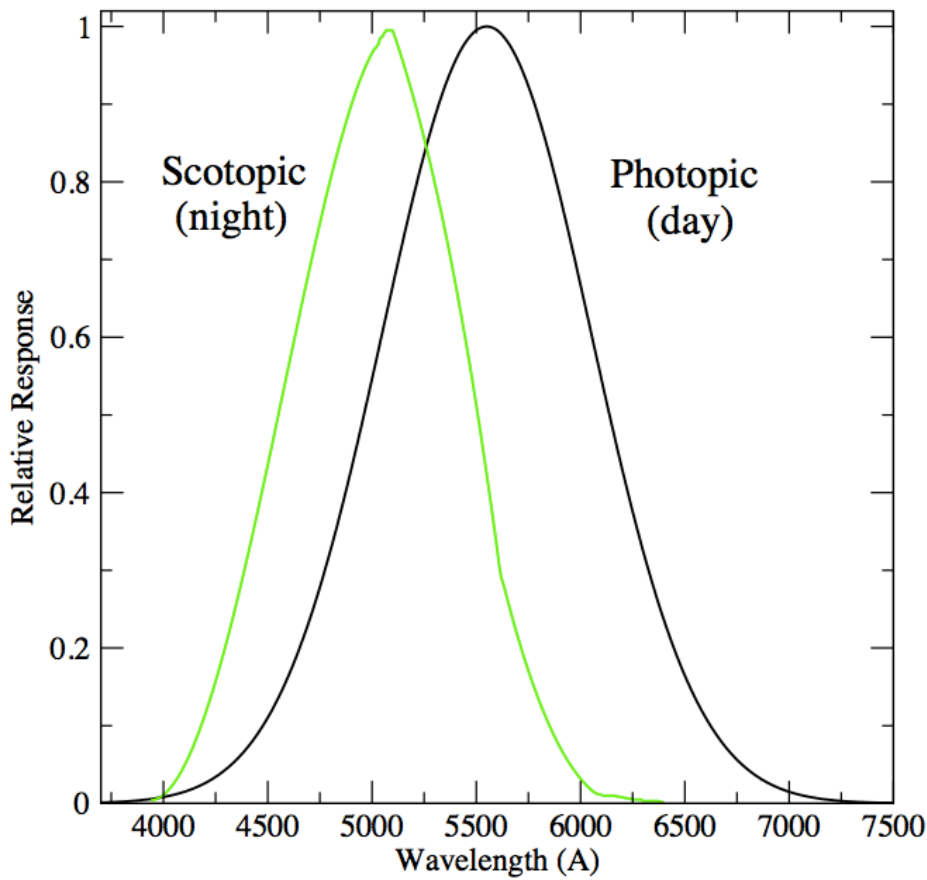

Fig. 5 – Relative response of the eye and retina for low light level conditions (scotopic curve) and for daytime (photopic) conditions.

The faintest stars visible to the unaided eye under dark sky conditions have apparent magnitude V ~ 6.0. We adopt this as a practical limit. (Author Krisciunas has reached apparent magnitude 6.3. Some keen-sighted observers, such as Brian Skiff[5] and Stephen O'Meara,[6] have proven that they can see stars fainter than V = 8.0.) A star of magnitude zero is $2.51189^6$ ~ 251.2 times brighter than a 6th magnitude star. One might think, then, that the limit of seeing a candle flame would be the square root of 251.2 (about 15.85) times 392 meters, or 6.2 km (3.86 miles). However, this neglects the fact that when we look at a bright star, there are enough photons to see the color of the star. We are using our phototopic vision, which uses the cones in our retina. When we look at the faintest star we can see, we use the rods in our retina, and scotopic vision applies.

In Fig. 6 we show the significance of using our daytime (photopic) vision vs. our nighttime (scotopic) vision on Vega and a candle flame. Because the spectral energy distribution and photon energy density fall off rapidly at the short wavelength end, the 480 A shift from phototopic to scotopic vision makes a significant difference in the results. To answer the question posed in our title, we need to determine how far a candle flame would have to be to appear of equal brightness to a 6th magnitude star with the spectral energy distribution of Vega. So we take the photon energy density of a 10,000 deg K black body for Vega, multiply it by the scotopic luminosity function of the eye, and integrate from 3700 to 10,000 A. We take the photon energy density of a 1400 deg K black body for the candle flame, multiply by the scotopic luminosity function of the eye, scale by 6.016 X $10^6$, and integrate from 3700 to 10,000 A. Now the ratio of flame function to star function is 0.2312. A mythical creature whose night vision luminosity function is the same as our day vision luminosity function would observe the candle flame to be comparable in brightness to Vega at 338 m times the square root of 0.2312, or about 162.5 m. A 6th magnitude star with the spectral energy distribution of Vega would be 15.85 times more distant, or 2576 m (roughly 1.60 miles). A candle flame situated at 10 miles (16093.5 m) would have an apparent brightness of V = 5 log (16093.5/162.5) ~ 9.98 mag. This is far beyond the capabilities of the most sensitive human eyes.

We have not used an astrophysically realistic spectral energy distribution for Vega. It has absorption lines primarily at the blue end. Perhaps a more realistic photospheric temperature to use for Vega is 9550 K, as adopted by Kurucz.[7] A candle flame spectrum is not a perfect black body spectrum either. There are some emission features. According to the Simbad database[8] Vega's V magnitude is 0.03, not 0.00, but this difference hardly matters, as the Johnson V-band is not the same as either luminosity function of the eye. For our purposes here many complications of alternate analysis would probably be swept under the rug with the calculation of a new scale factor to produce a signal from a candle flame at 338 m, which is 9.315 times stronger than the signal from Vega high in the sky and observed under clear sky conditions. Finally, if we were going to take this experiment one step farther by setting up a candle flame at a distance of 2.6 km to confirm our results, we would have to consider the horizontal atmospheric extinction along the line of sight to the candle flame, which would be problematic.

Nevertheless, we have shown that a candle flame at roughly 2.6 km would have an apparent brightness comparable to a 6th magnitude star. Could the keenest human eyes on the planet see a candle flame at 10 miles? We have provided strong evidence that the answer is No, for it would be as faint as a star of apparent magnitude 10, and that would require a pair of 7 X 50 binoculars mounted on a tripod, even for experienced observers with good night vision.

## Acknowledgments

We thank Bradley Schaefer for many useful discussions relevant to the proper interpretation of the CCD measurements.

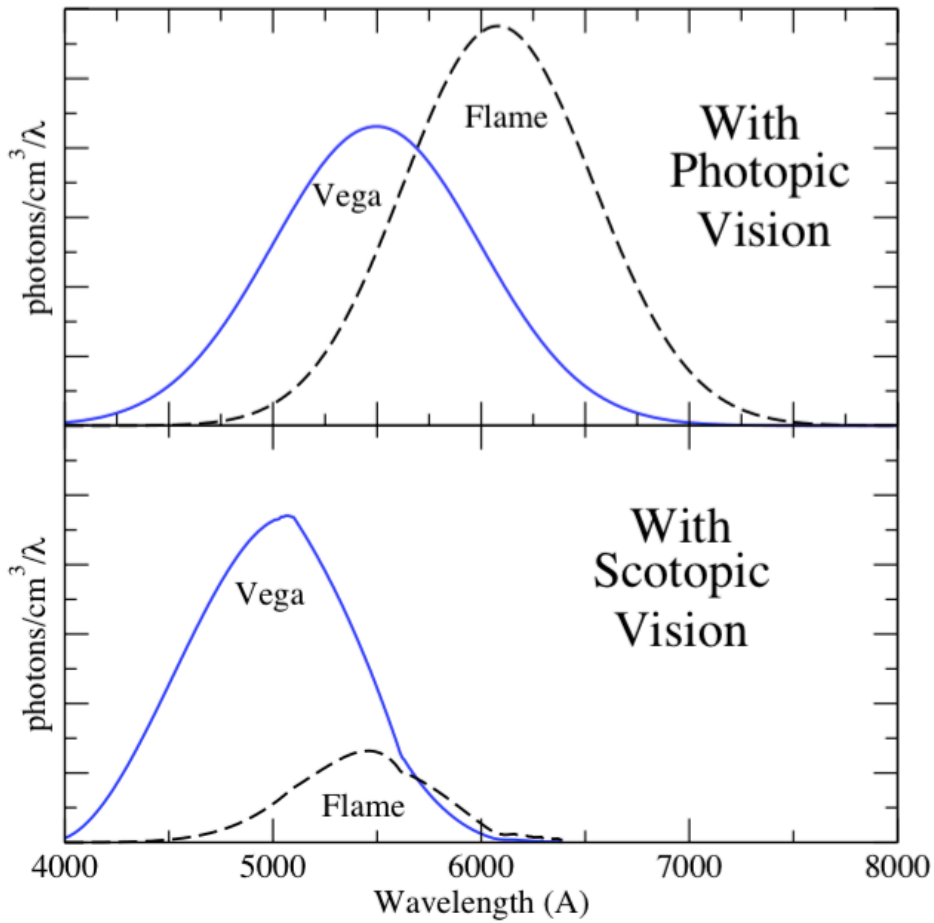

Fig. 6 – Top: relative brightness of candle flame at 338 m and Vega using daytime vision. Bottom: relative brightness of candle flame at 338 m and Vega using the night vision luminosity function of the eye. A star as bright as Vega is seen by our unaided eyes using day vision (cones).

# Footnotes

1. http://www.ispot.tv/ad/75pr/centrum-silver-your-eyes, accessed July 15, 2015.

2. IRAF is distributed by the National Optical Astronomy Observatory, which is operated by the Association of Universities for Research in Astronomy, Inc., under cooperative agreement with the National Science Foundation (NSF).

3. https://en.wikipedia.org/wiki/Planck%27s_law, accessed July 15, 2015.

4. https://en.wikipedia.org/wiki/Luminosity_function, accessed July 15, 2015.

5. Skiff, B., personal communication, 1999.

6. O'Meara, S., personal communication, 1992.

7. For synthetic photometry and calibration of supernova photometry author Krisciunas and his colleague N. Suntzeff have used for many years a Vega model spectrum by Kurucz obtained as file veg090250000p.asc5. The reader can generate model spectra for a variety of stars of different spectral types and compositions via the website
http://www.stsci.edu/hst/observatory/crds/castelli_kurucz_atlas.html

8. http://simbad.u-strasbg.fr/simbad/